\documentclass[fleqn,10pt]{wlscirep}
\usepackage[utf8]{inputenc}
\usepackage[T1]{fontenc}
\usepackage{bm}
\usepackage{wrapfig}
\usepackage{multicol}
\usepackage{parallel}
\usepackage{subfigure}
\usepackage{braket}
\usepackage{amsfonts}
\usepackage{amssymb}
\title{First Principles Calculations for Topological Quantum Materials}

\author[1]{Jiewen Xiao}
\author[1,*]{Binghai Yan}
\affil[1]{Department of Condensed Matter Physics, Weizmann Institute of Science, Rehovot 7610001, Israel}

\affil[*]{e-mail: binghai.yan@weizmann.ac.il}

\begin{abstract}
Discoveries of topological states and topological materials reshape our understanding of physics and materials over the last 15 years. 
First-principles calculations have been playing a significant role in bridging the theory of topology and experiments by predicting realistic topological materials. 
In this article, we overview the first-principles methodology on topological quantum materials. First, we unify different concepts of topological states in the same band inversion scenario. 
Then, we discuss the topology using first-principles band structures and newly-established topological materials databases. We stress challenges in characterizing symmetry-independent Weyl semimetals and calculating topological surface states, closing with an outlook on the exciting transport and optical phenomena induced by the topology.

\end{abstract}

\begin{document}

\flushbottom
\maketitle

\thispagestyle{empty}

\section*{ H1 Introduction}

Over the past decades, first-principles materials calculations have been playing a significant role in bridging fundamental theory and experiments. In particular, this has impacted the field of topological materials. For example, the last decade witnessed the successful prediction of the Bi$_2$Se$_3$-family of topological insulators (TIs)~\cite{zhang2009topological,xia2009observation}, the SnTe-class of topological crystalline insulators (TCIs)~\cite{hsieh2012topological} and the TaAs-family of Weyl semimetals (WSMs)~\cite{weng2015weyl,huang2015weyl}, stimulating the rapid development of this field. Assisted by the prediction power of first-principles calculations on weakly-interacting systems, theory usually leads the experimental research in topological materials, which is rare in the field of condensed-matter physics. 

Guided by the topological classifications, first-principles calculations identify topological materials by computing the bulk topological invariants (for example, the $Z_2$ index characterized by the parity for TIs \cite{fu2007topological1} ) and the topological surface states (TSSs). Then related materials can be synthesized and characterized usually by measuring TSSs via angle-resolved photoemsission spectroscopy \cite{yang2018visualizing,lv2019angle} or scanning tunneling microscopy \cite{avraham2018quasiparticle,zheng2018quasiparticle}. Topological band theory, first-principles calculations, materials preparation, and materials measurement cooperate intimately in a back-and-forth way as the materials design. In this Review, first-principles calculations mainly refer to the density-functional theory (DFT) \cite{kohn1999nobel} and DFT-derived effective models such as the localized Wannier functions \cite{Marzari2012}. Studies on topological systems also push today's first-principles calculations to pay more attention to the consequences of the time-reversal symmetry (TRS) and crystalline symmetries on the band structure than before.  

There are many excellent reviews on the general topological states, topological materials 
\cite{hasan2010colloquium,qi2011topological,Hasan2011,yan2012topological,bernevig2013topological,hosur2013recent,Witczak2014,Vafek2014,hasan2015topological,ando2015topological,chiu2016classification,ando2013topological,fang2016topological,yan2017topological,Hasan2017,Burkov2018,armitage2018weyl,shun2018topological}, and related first-principles calculations \cite{weng2014exploration,weng2016topological,Bansil2016,yu2017topological,hirayama2018topological,gao2019topological}.
This Technical Review aims to overview the state-of-the-art first-principles methodology and provide a beginner's guide in the study of topological materials. First, we heuristically unify different topological states in the same band inversion scenario and then demonstrate the TSSs by the bulk-boundary correspondence. 
Second, we introduce the topological invariants and symmetry indicators, which distinguishes a topological state from atomic insulators. 
Next, we present the general first-principles procedure to extract the bulk topology and compute TSSs by representative topological materials, including TIs\cite{kane2005z,kane2005quantum,bernevig2006quantum,konig2007quantum},
topological crystalline insulators (TCIs) \cite{fu2011topological}, Dirac semimetals (DSMs) \cite{young2012dirac,Wang2012}, Weyl semimetals (WSMs), \cite{wan2011topological,murakami2007phase,volovik2003universe} nodal line semimetals (NLSMs) \cite{Burkov2011,fang2015topological} and higher-order topological insulators (HOTIs) \cite{Zhang2013,song2017d,benalcazar2017quantized,Langbehn2017,schindler2018higher1}. We focus on weakly interacting topological materials in this Review as first-calculations are limited in their ability to handle strongly-correlated systems. In addition, we provide tutorial documents for all examples in supplementary information (SI). 

\section*{H1 Topological band theory}

Band theory was a seminal success in applying quantum mechanics to solid materials. After nearly a century, it is incredible to discover qualitatively new phenomena in this field. The TI is such a case, which is characterized by the robust metallic surface states inside the bulk energy gap. The TI band structure is intuitively characterized by a band inversion \cite{bernevig2006quantum} --- in which the valence and conduction bands switch order compared to an ordinary insulator. Up to our knowledge, William Shockley \cite{Shockley1939} first pointed out that band inversion leads to unique states on the surface by a spinless model. The Shockley surface states represent charge polarization, which is sensitive to changes in the surface potential. Further, a band inversion leads to the gapless bulk phase in the 2D or 3D spinless model without spin-orbital coupling (SOC). Thus, the Shockley state was debated and ignored as time passed (see REF.~\citeonline{martin_2020} for an overview). In 2006, band inversion and SOC were combined to propose a TI, which was then experimentally realised in HgTe/CdTe quantum wells \cite{bernevig2006quantum}. 
To be accurate, the band inversion refers to the order change between the valence and conduction bands with opposite parities~\cite{fu2007topological1}. In the absence of well-defined parities (such as inversion symmetry-breaking),
we consider the band inversion a heuristic scenario by assuming that the inversion-breaking bands are adiabatically connected to inversion symmetric states~\cite{qi2011topological,hasan2010colloquium}.
In the following discussions, we assume a 3D system with the TRS and include SOC unless otherwise specified. 

\subsection*{H2 Bulk-boundary correspondence and TIs}
\begin{figure}[h!]
    \centering
    \includegraphics[width=12cm]{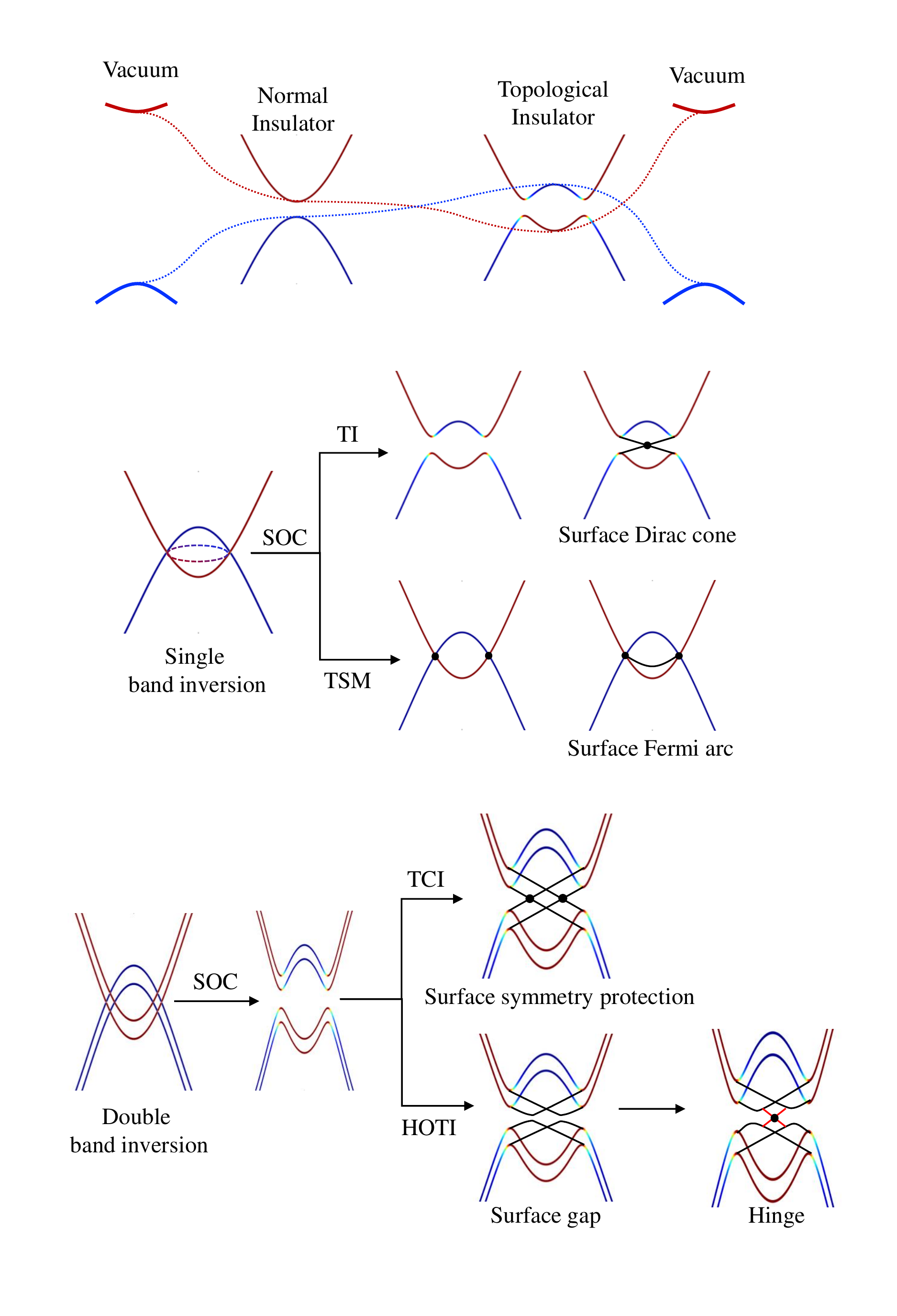}
    \caption{\textbf{Band inversion, gapless boundary states, and the bulk boundary correspondence}  Band inversion is well defined in the inversion symmetric system, but can still provide a heuristic scenario for the inversion-breaking case. \textbf{a} | The illustration of band inversion and metallic boundary states. On the interface between two insulators with opposite band orders, the band gap needs closing at the interface, exhibiting metallic surface states. Here the vacuum can be regarded a normal insulator. \textbf{b} | Single band inversion. The band inversion can form nodal ring (dashed ring) structure without spin-orbit coupling (SOC). If SOC gaps all the nodal ring, it leads to a topological insulator (TI) with Dirac-cone-type surface states (black lines). If some individual points at or near the nodal ring (filled black circles) remain degenerate, it leads to topological semimetals (TSM) such as the Weyl semimetal. On the surface of a Weyl semimetal, Fermi arcs (black line) exists to connect two Weyl points in the Fermi surface.  \textbf{c} | Double band inversions. If SOC leads to a whole gap, double band inversions induce two Dirac-type surface states. It can present a topological crystalline insulator (TCI) if there is a crystal symmetry (for example, the mirror symmetry) to protect two Dirac cones from gapping each other at special momenta. If two surface Dirac cones are gapped out, it may lead to a higher-order topological insulator (HOTI) when hinge states protected by the crystal symmetry exist on the boundary of two gapped surfaces. A TCI is insulating in the bulk and metallic on the symmetry-preserved surface, while a HOTI is insulating in both bulk and surface but metallic at some hinges.}
    \label{fig:BI}
\end{figure}

\begin{figure}[h!]
    \centering
    \includegraphics[width=\textwidth]{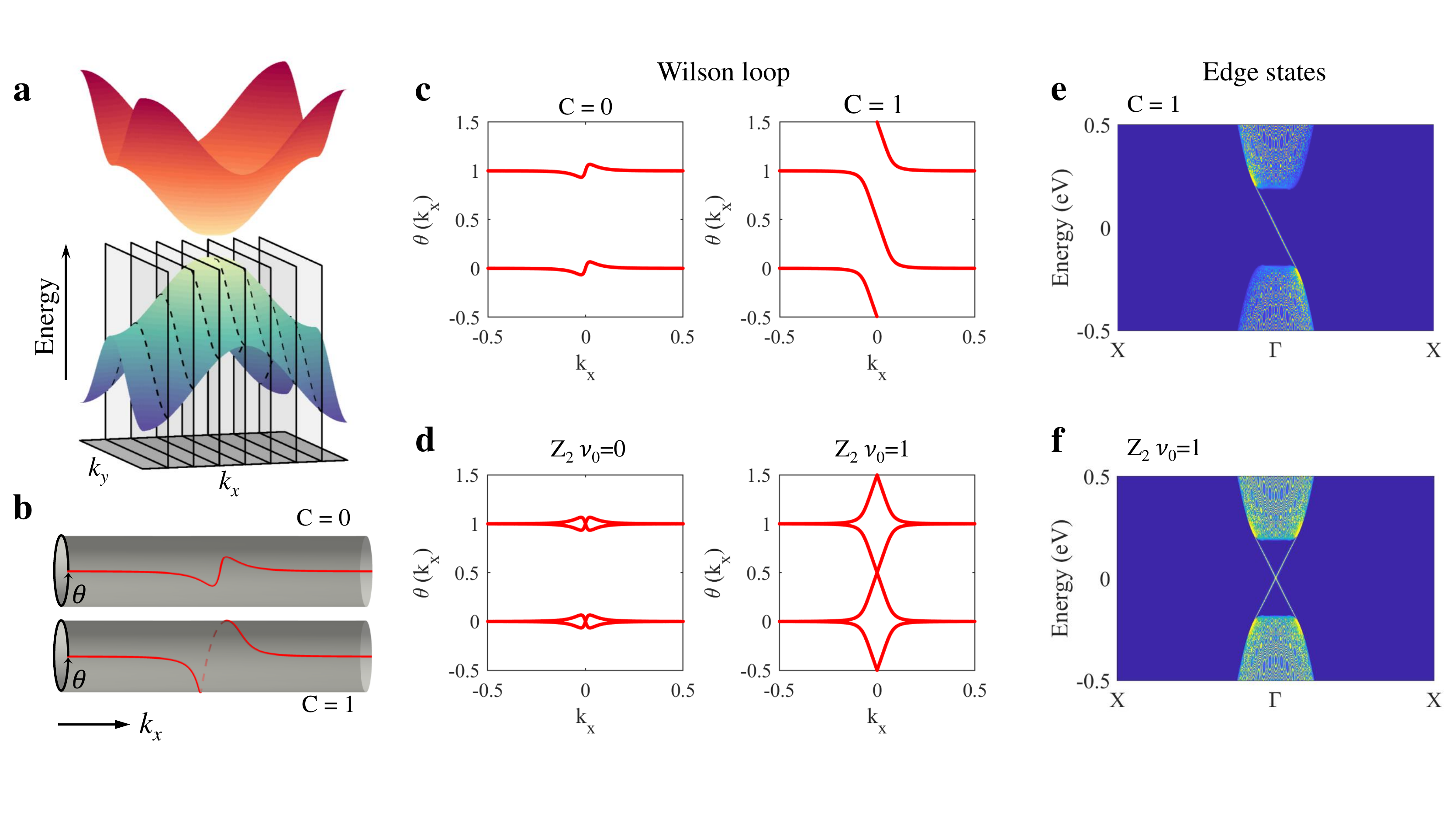}
    \caption{\textbf{2D Topological states and the Wilson loop.} \textbf{a} | Band structure of a magnetic insulator in the 2D Brillouin zone. The conduction and valence bands are separated by a gap. For generic $k_x$, the valence state disperses along $k_y$ periodically (dashed lines). We note $k_{x,y} \in$ [0,1] in the unit of reciprocal lattice rectors. Here, the Wilson loop or the Wannier charge center ($\theta(k_x)$) corresponds to the Berry phase (divided by $2\pi$) acquired along the $k_y \in $[0,1] line at given $k_x$ for the valence band. \textbf{b,c} | The Wannier charge center evolution (along $y$ axis) or the Wilson loop in the parallel transport. In the cylinder coordinate, one can find the trivial (Chern number $C=0$) and nontrivial ($C=1$) winding of $\theta(k_x)$. For the $C=1$ case, the charge pumping from one unit cell to another along the $y$ leads to the existence of edge state when the lattice is truncated along the $y$ direciton. \textbf{d} | The Wilson loop for a normal insulator ($\nu_0 =0$) and a 2D topological insulator (TI) ($\nu_0 =1$). For the TI, partners in the Kramers pair are pumped in opposite directions and form the Kramer pair in the next TRIM with new partners. \textbf{e,f} | Edge states exist on the boundary (along $k_x$) that is normal to the charge pumping direction ($y$). The edge state dispersion can be continuously deformed into corresponding Wilson loop spectra in \textbf{c,d}, demonstrating the bulk-boundary correspondence}
    \label{fig:2D}
\end{figure}

Consider a normal insulator and a TI, which exhibit opposite band orders at the center of the Brillouin zone (Fig.~\ref{fig:BI}). To move smoothly from one material to the other, the band gap needs to close at the interface because of the band inversion. Therefore, the interface exhibits gapless boundary states. 
We can naively classify all insulators into two classes, with and without the band inversion, and consider the existence of metallic states at the interface. A normal insulator is not necessarily metallic on the surface, for example, if it is interfaced with a vacuum (approximated to be a normal insulator too). In contrast, a TI is always metallic on its surfaces. Such an argument only requires  bulk band inversion and is robust against the surface perturbations such as defects or atomic reconstructions.
Therefore, the band inversion in bulk dictates the existence of metallic states, the TSSs, on the boundary, which is usually called the bulk-boundary correspondence~\cite{hatsugai1993edge,essin2011bulk}. 
The TI surface states usually form a single Dirac cone inside the bulk gap and connect the bulk valence and conduction bands. 
The Dirac point is protected by the TRS, which induces  double spin-degeneracy called the Kramers degeneracy. Therefore, it is not surprising that TRS also defines the $\mathbb{Z}_2$-type topological invariant of the bulk states. 
Furthermore, the bulk-boundary correspondence determines both the existence and the dispersion profile of TSSs.
The adiabatic charge pumping in the bulk, which can be evaluated by the Wannier charge center evolution or Wilson loop method ~\cite{yu2011equivalent,Soluyanov2011,alexandradinata2014wilson,weng2014exploration}, is actually a continuous deformation of the surface band structure~\cite{fidkowski2011model,taherinejad2014wannier}. We take a 2D insulator as an example in  Fig. \ref{fig:2D}.
Because of the periodicity, we can consider the $k_y$ line as a loop ($C_{k_y}$) in the Brillouin zone (from 0 to 1 in the unit of reciprocal lattice vector) for given $k_x$. 
For a single occupied band (Fig.~\ref{fig:2D}a), the Berry phase (divided by $2\pi$) acquired around the $C_{k_y}$ loop represents the Wannier charge center of the Bloch wave function, such that 
\begin{equation*}
\theta (k_x) = \frac{1}{2\pi} \oint_{C_{k_y}} dk_y A(k_x,k_y) 
\end{equation*}
where 
$A(k_x,k_y) = \braket{u(k_x,k_y)|\frac{\partial}{i\partial k_y}|u(k_x,k_y)} $ is the Berry connection. 
As $k_x$ varies from 0 to 1, the $\theta (k_x)$ profile corresponds to the charge pumping along the $y$ axis, which is called parallel transport of Wannier charge centers. 
In the case of many valence bands, the matrix representation of parallel transport is called a Wilson loop, and its eigenspectrum ($\theta (k_x)$) comprises the non-Abelian Berry phase factors.
If $\theta (k_x)$ shifts from one unit-cell to another in this process, it indicates a Chern insulator with Chern number $C=1$. 
One can also observe the nontrivial winding of $\theta (k_x)$ in the cylinder coordinate (Fig. \ref{fig:2D}b). 
The nontrivial charge pumping along $y$ indicates the existence of edge state when truncating the $y$ axis, because the pumping induces charge accumulation/depletion on the edge. The Wannier charge center evolution (Fig. \ref{fig:2D}c) is adiabatically connected to the corresponding edge band structure (Fig.. \ref{fig:2D}e). For a TRS-insulator, $\theta (k_x)$ includes Kramer pairs (Fig. \ref{fig:2D}d) because of TRS and are degenerate at time-reversal-invariant momenta (TRIM) $k_x = 0, 0.5$. For a normal insulator, the Kramers pair remain the same partners between neighboring TRIM. For a TI, however, the Kramers pair exchange their partners with the pair from another unit cells, displaying the same topology as their edge states (Fig. \ref{fig:2D}f). Such a correspondence between the bulk Wilson loop and boundary states will be demonstrated for different topological phases in the following sections.

\subsection*{H2 TCIs and HOTIs.}
\begin{figure}[h!]
    \centering
    \includegraphics[width=10cm]{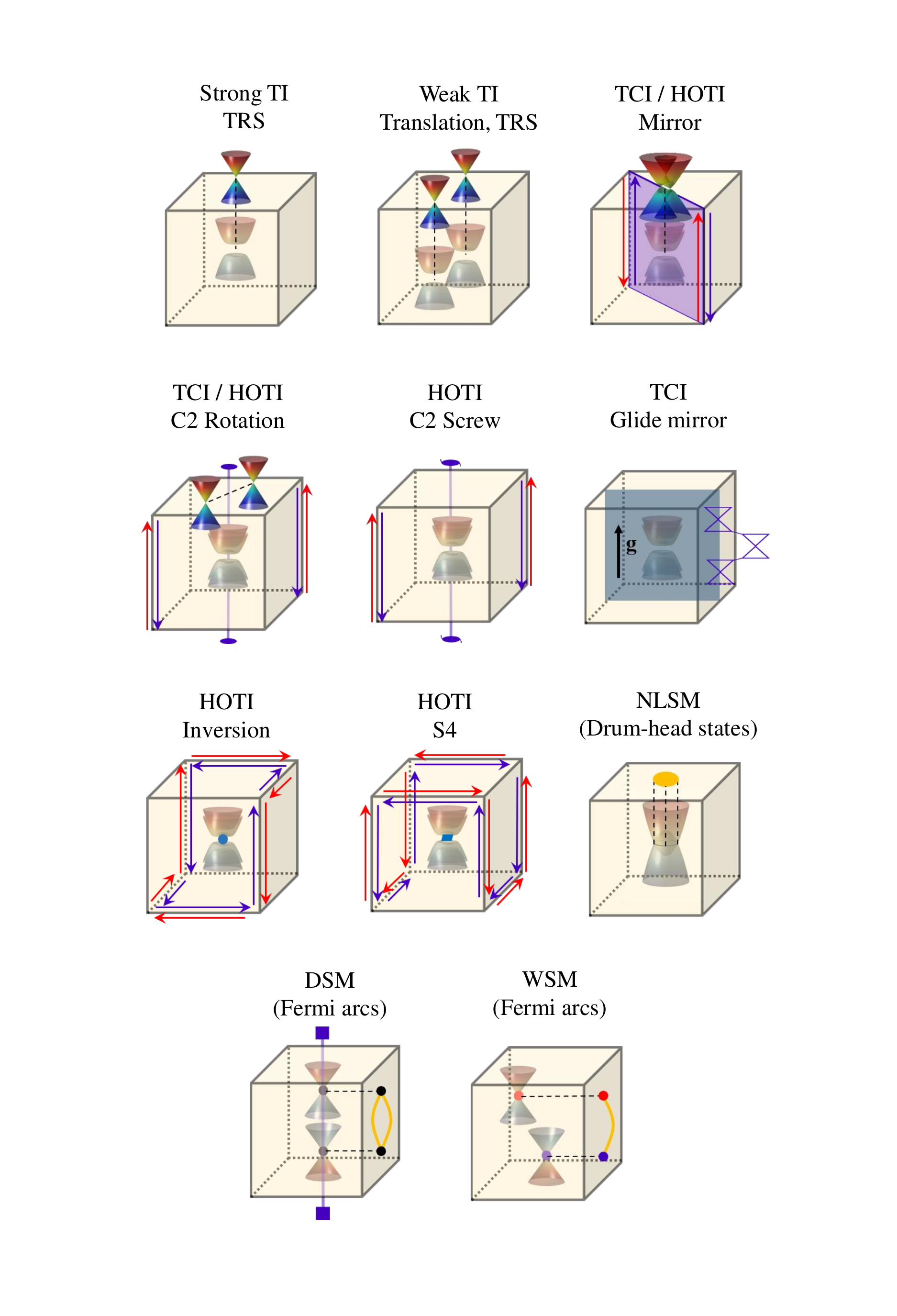}
      \caption{\textbf{Illustration of the bulk and surface states for different topological phases}. \textbf{a} | Strong topological insulator (TI). The single band inversion leads to single surface Dirac cones protected by the time-reversal symmetry (TRS). \textbf{b} | Weak TI. the double band inversion happens at different momenta, which leads to double surface Dirac cones on side surfaces protected by the translational symmetry and TRS. \textbf{c} | Mirror topological crystalline insulator (TCI) or mirror protected higher-order TI (HOTI). Double surface Dirac cones, which are induced by double band inversions, hybridize with each but do not gapped out on mirror-symmetric lines. When the nontrivial mirror cross the hinge, it leads to helical hinge states, (HOTI). \textbf{d} | $C_2$ rotation TCI. Double surface Dirac cones, as induced by double band inversions, exist on $C_2$ preserving surfaces. Two Dirac cones occur at generic momenta and related by $C_2$. Hinge states also exist and connect top and bottom metallic $C_2$ surfaces (rotation protected HOTI). \textbf{e} | $C_2$ screw HOTI. There are no surfaces that respect the $C_2$ screw symmetry, but two hinge states can exist and protected by $C_2$ screw. \textbf{e} | Hourglass TCI. Hourglass surface states exist on glide-symmetric lines. \textbf{f,g} | Inversion/$S_4$ HOTI: there are no surfaces that respect the inversion/$S_4$ symmetry but exist hinge states on inversion/$S_4$ symmetric lines. \textbf{h} | Nodal line semimetal (NLSM). The nodal line structure is usually formed by band inversion without SOC. Corresponding drumhead-like states occur on surfaces, but they are not protected by symmetries. \textbf{i} | Dirac semimetal (DSM). The band inversion induced DSM features two four-fold degenerate Dirac points on the rotation axis, which are rotation symmetry protected band crossings. A pair of corresponding Fermi arcs occur on surfaces. \textbf{j} | Weyl semimetal (WSM). The WSM features two-fold degenerate accidental-crossing points, called Weyl points, without any symmetry protection. The bulk Weyl points dictates surface Fermi arcs, which connect Weyl points with the opposite chirality (red and blue dots). }
    \label{fig:BI2}
\end{figure}

The bulk-boundary correspondence indicates that one band inversion leads to one surface Dirac cone when projecting the inverted bulk bands to the surface. If an odd number of band inversions occur, the TRS can protect the Dirac states. If an even number of band inversions exist, only the TRS is not enough to preserve them. We may need the crystal symmetries or their combined symmetry with TRS for protection, leading to the topological crystalline insulators (TCIs).  
Take a mirror-protected TCI for example (Fig.~\ref{fig:BI2}). 
Double band inversions give rise to double surface Dirac cones that cross with each other. 
Two Dirac cones are thus hybridized and gapped out in generic $k$-points except two points (black dots in Fig.~\ref{fig:BI}) in the mirror plane. Only the surfaces that preserve the mirror symmetry are still metallic, while surfaces without mirror symmetry are insulating. 
When surfaces do not preserve the mirror symmetry, the hinge between two such surfaces could be still mirror-symmetric. And on the hinge one can expect protected hinge states (see FIG.~\ref{fig:BI2}).
This is an example of the mirror-protected HOTI \cite{hsieh2012topological}. The bulk and surface states for TCIs and HOTIs are illustrated in Fig.~\ref{fig:BI2}, which is closely related to the crystal symmetry. Generally, there are seven kinds of topological invariants associated with crystal symmetries, including translation, mirror, glide, rotation, screw, inversion, and $S_4$ in 230 space groups (SGs) \cite{song2018quantitative,khalaf2018symmetry}, resulting in the diverse TCI/HOTI phases in Fig.~\ref{fig:BI2}.  

\subsection*{H2 Topological semimetals}
The inverted valence and conduction bands are usually degenerate at the crossing points if SOC is turned off or negligibly small. These degenerate points are called nodal points. If they form lines or rings, they are referred to as nodal lines or nodal rings (Fig.~\ref{fig:BI}b).
If SOC gaps all crossing points, we obtain topologically insulating phases discussed above. Even with SOC, some individual points or lines may remain degenerate. 
Thus, topological semimetals refer to the gapless bulk states with SOC and mainly include the DSM, WSM, and NLSM. Because of the lack of a bulk gap to isolate the valence bands, topological invariants for topological semimetals are not well defined here.
However, we can still expect special surface states, because topological semimetals usually originate from the band inversion.
Especially for a WSM, TSSs form open arcs connecting Weyl points with opposite chirality on the Fermi surface\cite{wan2011topological}, different from closed Fermi loops or extended Fermi lines for any other 2D states, as illustrated in Fig.~\ref{fig:BI2}. 

\section*{ H1 Symmetry indicators}\label{section-SI}
The band inversion is a heuristic scenario to understand bulk topology and the corresponding TSSs. To identify the topological phase accurately, 
we need the information of the wave functions in the momentum space. 
Compared to the direct calculation of topological invariants, the symmetry indicator \cite{po2017symmetry,bradlyn2017topological,bradlyn2018band,kruthoff2017topological} provides a convenient way to probe the band topology from the wave function symmetry. 
The symmetry indicator only requires the information of the band representations \cite{zak1981band,Zak1982,Bacry1988,dresselhaus2007group} at high-symmetry momenta and can be easily implemented into first-principles calculations for real materials \cite{vergniory2019complete,tang2019comprehensive,zhang2019catalogue}. 
Such a symmetry approach was epitomized by the Fu-Kane parity criterion \cite{fu2007topological1} to determine TIs with the inversion eigenvalues and recently generalized to all space group symmetries \cite{po2017symmetry,bradlyn2017topological,bradlyn2018band,kruthoff2017topological}.
The insight behind the symmetry indicator is that occupied states of the topological phase cannot be represented by symmetrical localized Wannier orbitals \cite{Soluyanov2011}, distinct from atomic insulators (AIs). 
The band representation, which is based on the wave function symmetry, provides a tool to extract the symmetry indicator and identify the equivalence or distinction between a band insulator and AIs.

Generally, the symmetry data for any given band structure (occupied bands) is written as a $\boldsymbol{n}$ vector (not a true vector but has the linear structure)~\cite{tang2019efficient}. Its elements include the number of valence bands $\nu$ and the number of each irreducible representation (irrep) at each high-symmetry point. These elements are constrained by the fixed electron filling and the high-symmetry lines in each space group (SG), known as the compatibility relation \cite{michel1999connectivity,aroyo2006bilbao1,bradley2009mathematical,po2017symmetry,bradlyn2017topological,kruthoff2017topological}. 
For the gapless case, the $\boldsymbol{n}$ vector violates the compatibility relation, which indicates the existence of topological semimetals with band crossings at high-symmetry points/lines. 
For the gapped case (with no band crossing at high-symmetry points/lines), all the $\boldsymbol{n}$ vectors satisfy the linear compatibility relations and form a linear space, denoted as $\{BS\}$, which includes both gapped trivial and nontrivial symmetry entries. 
On the other hand, to construct the space of AIs, a set of localized symmetric Wannier orbitals are placed at Wyckoff sites, which form a set of AIs \cite{tang2019efficient,po2020symmetry}. With the real space description, the corresponding $\boldsymbol{n}$ vectors in the momentum space are retrieved by the Fourier transform. They also form a linear space, denoted as $\{AI\}$ as the $\{BS\}$ for an atomic insulator, containing all the trivial symmetry entries. 
When subtracting $\{AI\}$ from the general $\{BS\}$, by solving the quotient group $\{BS\}/\{AI\}$ mathematically, it leads to the symmetry indicator group. 

The symmetry indicator group is usually generated by several $\mathbb{Z}_{n=2,3,4,6,8,12}$ and describes the mismatch of band representations between topological insulators and AIs. The nontrivial symmetry indicator value indicates the topology. For a given band structure, we can always express the $\boldsymbol{n}$ vector by a linear superposition of {AI} basis~\cite{tang2019efficient}. If the coefficients are integral (fractional), the band structure is trivial (nontrivial), and the symmetry indicator value can be derived based on the coefficients. We provide an example of the $p2$ wallpaper group in the supplementary information (SI) to illustrate the above concepts and procedures. 
Since the symmetry indicator only relies on the band representation at high-symmetry points, the symmetry indicator can be calculated in a similar way like the Fu-Kane parity criterion \cite{song2018quantitative,khalaf2018symmetry}. For the 230 SGs, formalism is derived to map symmetry indicators to topological invariants. 

Take SG $\#$2 with inversion symmetry as an example.
The Fu-Kane invariants ($\nu_0;\nu_1 \nu_2 \nu_3$) are determined by the inversion eigenvalues $\delta_{\boldsymbol{K}}^i$ of all occupied Kramers pairs $i$ at TRIM ${\boldsymbol{K}}$, which is $\pm 1$. For example, the strong TI index $\nu_0$
is determined by 

\begin{equation*}
(-1)^{\nu_0} = \prod\limits_{\boldsymbol{K},i} \delta_{\boldsymbol{K}}^i. 
\end{equation*}
In the symmetry indicator scheme, $\{BS\}$ is expressed by ($n_{\boldsymbol{K}}^{+},n_{\boldsymbol{K}}^{-},\nu$), where $n_{\boldsymbol{K}}^{+}$($n_{\boldsymbol{K}}^{-}$) represents the number of occupied even(odd)-parity states at $\boldsymbol{K}$. $n_{\boldsymbol{K}}^{\pm}$ is even because of the Kramers degeneracy. 
The corresponding symmetry indicator group is $\mathbb{Z}_2^3 \times \mathbb{Z}_4$. 
The $\mathbb{Z}_4$ type indicator is
\begin{equation*}
z_4=\sum\limits_{\boldsymbol{K}} \frac{n_{\boldsymbol{K}}^{-}-n_{\boldsymbol{K}}^{+}}{4} \rm{mod} 4.
\end{equation*}
One can recognize $\nu_0 = z_4$ mod 2, by 
\begin{equation*}
(-1)^{\nu_0} = (-1)^{\frac{1}{2} \sum\limits_{\boldsymbol{K}} n_{\boldsymbol{K}}^{-}} 
= (-1)^{\frac{1}{2} (4\nu +\sum\limits_{\boldsymbol{K}} 
\frac{n_{\boldsymbol{K}}^{-}-n_{\boldsymbol{K}}^{+}}{2})} = (-1)^{\sum\limits_{\boldsymbol{K}}\frac{n_{\boldsymbol{K}}^{-}-n_{\boldsymbol{K}}^{+}}{4}}
\end{equation*}
where $\nu = n_{\boldsymbol{K}}^{-}+n_{\boldsymbol{K}}^{+}$ is an even constant at all 8 $\boldsymbol{K}$ points. Because it is mod 4, $z_4$ contains more information than $\nu_0$. For instance, $z_4=2$ indicates the existence of a TCI.
Three $\mathbb{Z}_2$ type indicators ($z_{2}^j, j=1,2,3$) correspond to three weak invariants ($\nu_{1,2,3}$)  and are expressed as
\begin{equation*}
z_{2}^j=\frac{1}{2}\sum\limits_{\boldsymbol{K}}^{k_j = \frac{1}{2}} n_{\boldsymbol{K}}^{-} \rm{mod} 2,
\end{equation*}
where $k_j$ is in unit of reciprocal lattice vectors.

Technically, some useful tools (see TABLE I) can analyze the wave functions given by DFT, generate band representations, and compute symmetry indicators \cite{he2019symtopo,vergniory2019complete,gao2021irvsp}.
Based on these methods, topological materials databases have been established, giving the symmetry indicator classification for known solid materials.
For instance, SnTe (SG $\#$225) has the symmetry indicator group $\mathbb{Z}_8$ and the indicator is $z_8 = 4$, suggesting a TI/TCI. Except for SG $\#$174 and $\#$187-190, the largest order indicator, like the $\mathbb{Z}_{8}$ type indicator in SG $\#$225, and $\mathbb{Z}_{4}$ type in SG $\#$2, is called the strong factor \cite{song2018quantitative,khalaf2018symmetry}. The odd value of the strong factor indicates a strong TI, while the even value corresponds to a TCI. Therefore, SnTe is a TCI.
As shown in SI, the $z_8=4$ in SG $\#$225 corresponds to two possible collections of invariants, because symmetry indicators alone cannot fully determine what kind of TCI it belongs to. 
One set of invariants has the nontrivial mirror Chern number~\cite{teo2008surface,hsieh2012topological} $C_m=2$ on the (110) plane while the other set does not.
By calculating the mirror Chern number, a mirror Chern insulator (the former set) is recognized for SnTe.
Another example is Bi with SG $\#$166. The symmetry indicator group is $\mathbb{Z}_2^3 \times \mathbb{Z}_4$, and the indicator is (0002), belonging to a TCI. The (0002) value corresponds to two collections of invariants. By calculating the mirror Chern number of the $(\overline{2}10)$ plane, the nontrivial inversion, rotation $2^{(001)}$ and screw $2^{(001)}_1$ invariant can be identified. 
Generally, we can probe the band topology by calculating symmetry indicators and map them to a collection of invariants. Although such mapping is one to many, some specific invariant such as the mirror Chern number can help to identify the correct collection. With invariants, surface and hinge states can be anticipated according to the bulk-boundary correspondence and revealed by first-principles calculations.

The present symmetry indicator theories and databases commonly work for nonmagnetic topological materials with even electron fillings, and cannot capture the band topology at generic momenta. 
Thus, symmetry indicators usually cannot describe the WSM phase, although it indeed identifies WSMs in some special cases like WSMs with the inversion symmetry~\cite{hughes2011inversion,turner2012quantized} or the $S_4$ symmetry~\cite{gao2020high}.
Further, symmetry indicators, by definition, are indicators rather than topological invariants. Even if the symmetry indicator is trivial, the system may possess hidden topology that cannot be discerned by band representations, for example, a TI in SG $\#1$ with no extra symmetry except the identity.
Additionally, topological classifications of magnetic SGs are developed~\cite{watanabe2018structure,Elcoro2020}, and a preliminary materials database is built very recently~\cite{xu2020high1}.

\section*{H1 Topological Materials}
\subsection*{H2 Topological Insulators}\label{section:TI}
\begin{figure}[h!]
    \centering
    \includegraphics[width=\textwidth]{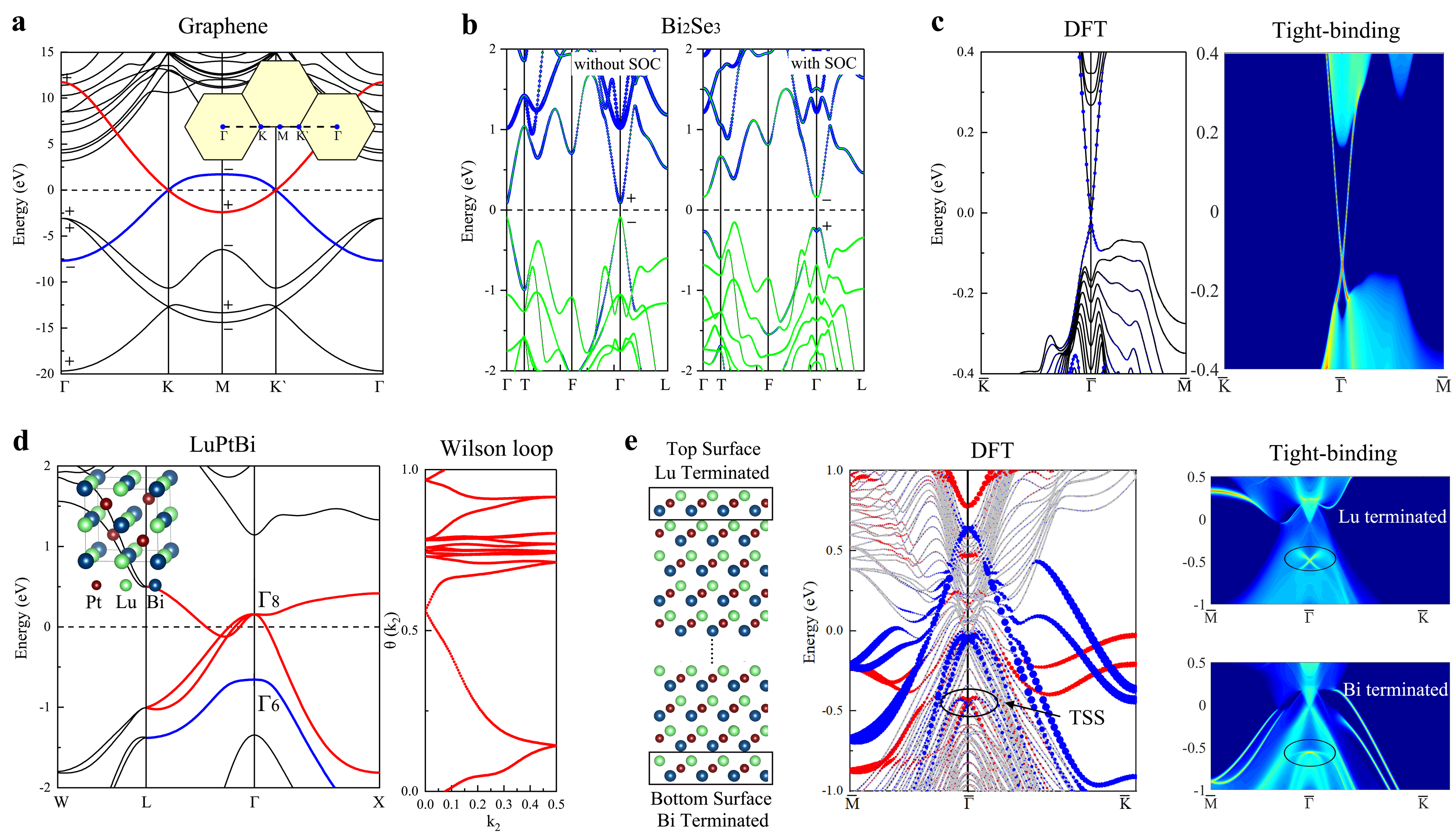}
      \caption{\textbf{Topological Insulators (TIs).} \textbf{a} | The band structure of the 2D TI, graphene, where the red and blue bands are inverted between $\Gamma$ and M. \textbf{b} | The band structure of the 3D TI Bi$_2$Se$_3$ with and without spin-orbit coupling (SOC). The blue/green dots are weighted by the projected $p$ orbitals of Bi/Se. The parity of the top valence band is inverted at $\Gamma$ when switching on SOC. \textbf{c} | The $(111)$ surface band structures from the first-principles density functional theory (DFT) calculations and tight-binding calculations. \textbf{d} | The band structure and the Wilson loop of LuBiPt, where the red and blue bands are inverted $\Gamma_8$ and $\Gamma_6$ bands. The Wilson loop only takes $\Gamma_6$ bands and below as valence bands and Wannier charge center $\theta$ evolves from $\Gamma$ to $L$. \textbf{e} | Schematic representation of the slab model for the $(111)$ surface and surface band structures from the first-principles DFT calculations and tight-binding calculations.}
    \label{fig:TI}
\end{figure}

Band structures of various TIs are shown in Fig.~\ref{fig:TI}.
To judge a TI by the band inversion, it is useful to compare band structures with and without SOC. For the presence of SOC, the $\mathbb{Z}_2$ topological invariant can be defined based on the TRS operator $\Theta^2=-1$ of the spinful fermion~\cite{kane2005z}. 
Without SOC, the TRS operator becomes $\Theta^2=1$ in the spinless case, and we only get either trivial insulators or semimetals/metals.
For example, Bi$_2$Se$_3$ is a trivial insulator without SOC and exhibits a band inversion when the SOC is tuned from zero to the experimental value, representing a topological phase transition from an ordinary insulator to a TI~\cite{zhang2009topological}. In the state-of-art DFT packages, the SOC amplitude can be easily tuned by a scaling parameter\cite{Yan2010}. In addition, strain can also be applied to examine the band inversion and topological phase transition in calculations. 

To identify a TI accurately, the parity criteria \cite{fu2007topological1}, based on the first-principles Bloch wave functions, is extensively used for insulators with the inversion symmetry. 
For an AI described by one symmetrical localized orbital with no dispersion, Bloch wave functions exhibit the same parity at all TRIM, leading to  parity product $(-1)^{\nu_0}=+1$. 
Other AIs are adiabatically connected to this extreme case, although parities at different TRIM may differ trivially.
If the band inversion occurs once by switching the conduction and valence band at some TRIM, the parity product switches from $+1$ to $-1$, generating a TI with $\nu_0=1$. 

The Bi$_2$Se$_3$ family of TIs is a well-known example to demonstrate the parity criteria. We note that graphene can also be recognized as a TI (including SOC) from the band inversion and parity. The TRIM in the 2D Brillouin zone are $\Gamma$ and $M$ points. Although it was extensively studied with the Dirac model near the $K$-point, between the $\Gamma$ and $M$ points (Fig.~\ref{fig:TI}a), we can find a clear order switching of two bands with opposite parities, also resulting in a TI. In addition, such a band inversion generates the Dirac cone at $K$ in the absence of SOC. 

The TSS of a TI is characterized by an odd number of Dirac cones at the surface TRIM. The TSS calculation is usually based on two schemes: first-principles DFT calculations and tight-binding calculations. First-principles DFT calculations simulate the surface with a slab model, which is compatible with the periodic boundary condition. The slab model should be thick enough to suppress hybridization between the top and bottom surfaces, which increases the computational cost dramatically. 
Tight-binding calculations, which usually use Wannier function parameters extracted from DFT bulk calculations, can efficiently calculate the half-infinite surface with the iterative Green's functions~\cite{sancho1984quick,dai2008helical,zhang2009electronic}. 
However, the first-principles calculations can capture the surface atomic relaxation and charge redistribution, and are usually more realistic to compare with experiments. Regardless, the surface states from different methods should provide the same topology. For instance, the surface Dirac cones from two methods are almost identical for Bi$_2$Se$_3$ (Fig.~\ref{fig:TI}c) because Bi$_2$Se$_3$ exhibits the smooth van der Waals type termination with  negligible surface charge redistribution compared to the bulk.

Materials without inversion symmetry and without van der Waals layered structures are more complicated. We take the half-Heusler compound LuPtBi \cite{chadov2010tunable,Lin2010} for example. As shown in the inset of Fig.~\ref{fig:TI}d, Pt and Bi form a zincblende lattice, and Lu fills the center position of the zincblende cube. Thus, LuPtBi exhibits strong chemical bonds and breaks the inversion symmetry. Around the $\Gamma$ point, there is an inversion between the $\Gamma_8$ and $\Gamma_6$ bands (similar to the case of HgTe \cite{bernevig2006quantum,fu2007topological1}). Because $\Gamma_8$ is four-fold degenerate, the band structure is gapless (even with SOC) at the Fermi level. 
However, there is a direct energy gap above $\Gamma_6$, which is relevant to the band inversion. This gap enables us to use a more general method, the Wannier charge center evolution or Wilson loop \cite{yu2011equivalent,Soluyanov2011,alexandradinata2014wilson,weng2014exploration}, to determine the topology. The Wilson loop method traces the Wannier charge center (Berry phase) evolution of valence wave functions between TRIM. 

Fig.~\ref{fig:TI}d presents the Wilson loop when taking $\Gamma_6$ band and below as valance bands. From $\Gamma$ to $L$ ($k_2$ from 0 to $0.5$), the Wannier centers $\theta$ with respect to the integration loop $k_1$ from 0 ($\Gamma$) to $1$ ($L$) exhibit a nontrivial pattern like Fig.~\ref{fig:2D}d(left), where $k_{1,2}$ are reciprocal lattice vectors. 
At $\Gamma$ and $L$, the Wannier centers are doubly degenerate as Kramers pairs. However, Wannier centers change their Kramers partners between $\Gamma$ and $L$, as a consequence of the band inversion. 
For a trivial insulator, Wannier centers still recombine with the same partners between TRIM (Fig.~\ref{fig:2D}d).
The Wilson loop also has the same topology as the dispersion of TSSs~\cite{taherinejad2014wannier}. Results in Fig.~\ref{fig:TI}d thus dictates the existence of Dirac surface states inside the gap between $\Gamma_6$ and $\Gamma_8$. 

To calculate the TSS of LuPtBi, the (111) surface is cleaved and has two possible terminations by Lu or Bi atoms. Both DFT and tight-binding calculations demonstrate the Dirac-cone-type TSSs at --0.5 eV in the inverted gap for both terminations. The tight-binding results are easier to recognize the topology since they ignore the trivial states. 
However, when comparing with experiments, first-principles results are more reliable to reveal the detailed dispersion of TSSs and other surface states~\cite{liu2016observation}.
The example of LuPtBi thus indicates that metallic bulk materials can also exhibit the nontrivial topology, and TSSs may appear far from the charge neutral point. For instance, TSSs were recognized on the surfaces of noble metals (also known as Shockley states)~\cite{Yan2015}, grey arsenic~\cite{Zhang2017As}, and in the empty states of BaBiO$_3$~\cite{yan2013large}.

\subsection*{H2 Topological Crystalline Insulators}

TCIs, including HOTIs, refer to topological phases with even band inversions protected by crystalline symmetries. The bulk topology only features topological surface and/or hinge states on suitable terminations. We assume the TRS in the following discussions, although some TCIs and HOTIs break TRS.
The first theoretically-predicted TCI is SnTe, protected by the mirror symmetry \cite{hsieh2012topological}. Driven by SOC, band inversions happen four times at four $L$ points, where the Sn-$p$ and Te-$p$ states are inverted (Fig.~\ref{fig:TCI}a). 
The $(110)$ and each equivalent mirror plane host two $L$ points and the double band inversion inside gives the nontrivial mirror Chern number $C_m = 2$. $C_m$ is defined as $C_m  = (C_{+i}-C_{-i})/2$, where $C_{\pm i}$ is the Chern number of the mirror eigenstate with eigenvalue $\pm i$ \cite{teo2008surface,hsieh2012topological}. 
The mirror protection can be intuitively interpreted as two copies of Chern insulators in the $\pm i$ subspace of the mirror, whose surface states can never be hybridized.
Furthermore, $C_m = 2$ means two pairs of helical edge states on the boundary of (110) mirror planes.
In contrast, (001) mirror planes are trivial with $C_m = 0$, since there is no band inversion inside this plane.

The Wilson loop method can demonstrate the nontrivial mirror Chern number. In both (110) and ($1 \bar{1}0$) mirror planes, Wilson loop shows two chiral modes, which are upwards and downwards flows for $\theta(k)$ along $k$ (Fig. \ref{fig:TCI}b), in each mirror eigenstate subspace. 
Therefore, it means $C_{+i}=-C_{-i}=\pm 2$ and $C_m = 2$. 
In the language of symmetry indicators, SnTe has the indicator $z_8=4$, which is mapped into the first collection of invariants shown in the SI. $z_8=4$ also gives $C_m=2$ for (110) mirror planes and other compatible invariants associated with the rotation and screw symmetry. 

Because of the bulk topology, the TSS on suitable terminations can be expected for SnTe. For the (001) surface, it is normal to $(110)$ and $(1 \Bar{1}0)$ mirror planes, which are projected into $\Bar{\Gamma}$-$\Bar{X_1}$ and $\Bar{\Gamma}$-$\Bar{X_2}$ in the surface Brillouin zone (Fig.~\ref{fig:TCI}a). For the (110) mirror, two L points ($L_2$ and $L_3$) inside are projected into same $\Bar{X_1}$ point. Thus, two surface Dirac cones are hybridized with each other and only remain gapless along the (110) mirror projected line. Such surface topology is reflected in the (110) Wilson loop: the [001] integration loop corresponds to the (001) surface termination and the dispersion along $[1\Bar{1}0]$ (i.e., $\Bar{\Gamma}$-$\Bar{X_1}$-$\Bar{\Gamma}$ in the surface Brillouin Zone) shares the similar topology with the TSS. 
Overall, the (001) surface hosts four Dirac cones on the four equivalent $\Bar{\Gamma}$-$\Bar{X}$ lines. Since Dirac cones are related by $C_4$, this scenario further satisfies the nontrivial [001] $C_4$ rotation invariant. Generally, the surface states of each invariant are compatible with each other.
Similarly, the (111) surface is normal to three mirror planes $(1\Bar{1}0)$, $(10\Bar{1})$ and $(01\Bar{1})$, which are projected into three equivalent $\Bar{\Gamma}$-$\Bar{M}$ lines. 
For each mirror plane, two L points inside are projected into $\Bar{\Gamma}$ and $\Bar{M}$ (Fig.~\ref{fig:TCI}a). 
Therefore, the scenario for the (111) surface is four Dirac cones sitting at one $\Bar{\Gamma}$ and three $\Bar{M}$ points, respectively.
This is consistent with the bulk $(1\Bar{1}0)$ Wilson loop. If the integration loop is along [111] and the acquired Berry phase disperses in the $[11\Bar{2}]$ direction (which is the $\Bar{\Gamma}$-$\Bar{M}$-$\Bar{\Gamma}$ line in the surface Brillouin Zone), the Wilson loop can correspond to the (111) TSSs (Fig.~\ref{fig:TCI}b).

\subsection*{H2 Higher order topological insulators}
\begin{figure}[h!]
    \centering
    \includegraphics[width=\textwidth]{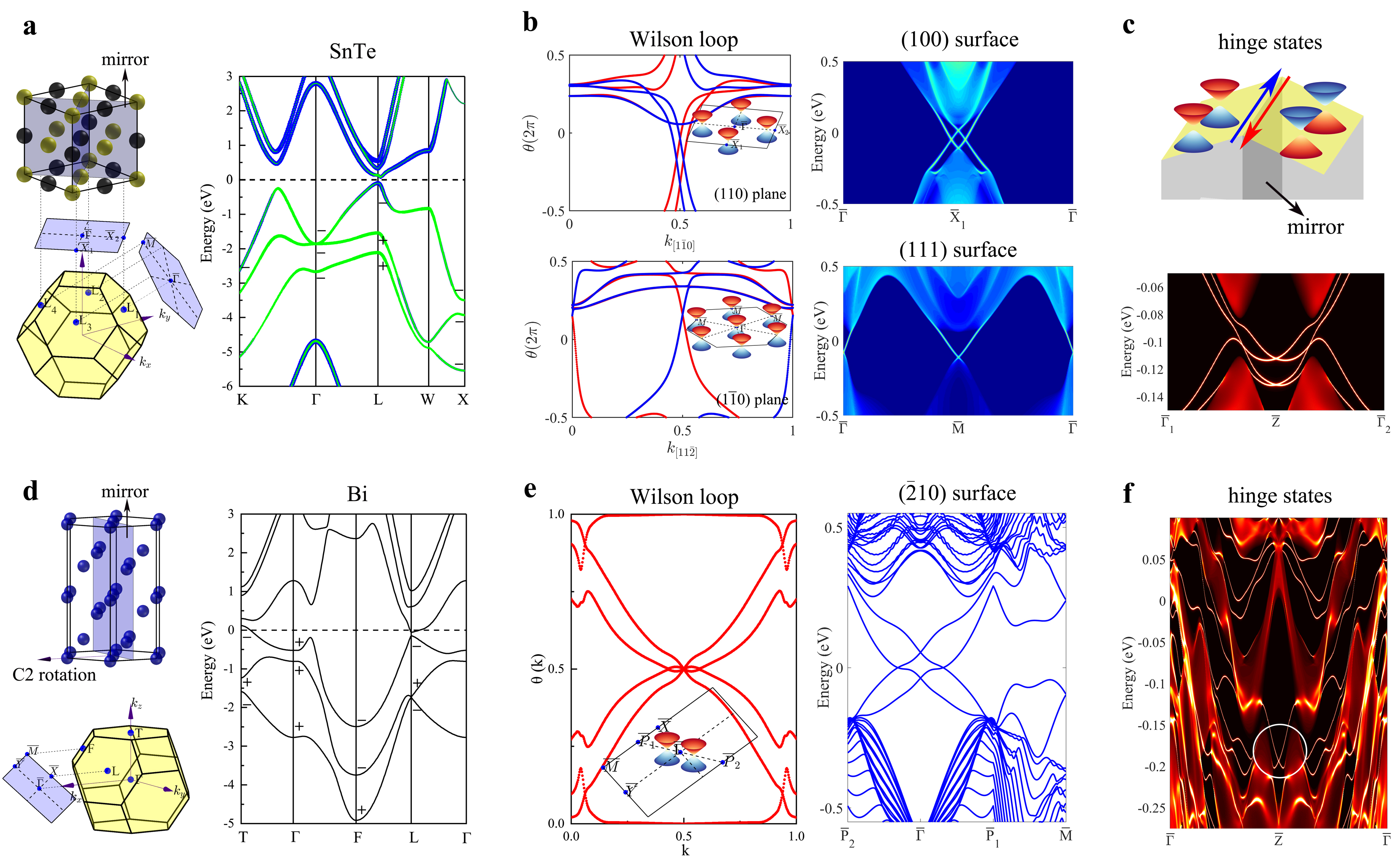}
     \caption{\textbf{Topological Crystalline Insulators (TCIs) and Higher-Order Topological Insulators (HOTIs).} \textbf{a} | The SnTe crystal structure and Brillouin zone. The (001) and (111) surface Brillouin zone is plotted, which corresponds to surface band structures in \textbf{b}. $\bar{\Gamma}-\bar{X_1}$ on (001) and $\bar{\Gamma}-\bar{M}$ on (111) surfaces are the projection of bulk $(110)$ and $(1\bar{1}0)$ mirror planes, respectively. The band structure of SnTe is presented on the right, where the blue and green dots are weighted by the projected $p$ orbitals of Sn and Te. It shows the band inversion happens at the $L$ point. \textbf{b} | The Wilson loop of (110) and $(1\Bar{1}0)$ mirror planes and the corresponding (001) and (111) surface states. The mirror resolved Wilson loops for $\pm i$ mirror eigenstates are presented by red and blue dots, respectively. Schematic representation of surface Dirac cones is shown in the inset. For the upper panel, the Wilson loop is integrated along $[001]$ and evolves along the $[1\bar{1}0]$ direction, which corresponds to the $(001)$ surface band structure along $\bar{\Gamma}-\bar{X_1}$. For the lower panel, the Wilson loop is integrated along $[111]$ and evolves along the $[11\bar{2}]$ direction, which corresponds to the $(111)$ surface band structure along $\bar{\Gamma}-\bar{M}$. \textbf{c} | Schematic representation of hinge states for the mirror TCI with TRS, where the red and blue arrows represent the Kramers pair. The gapped Dirac cones on two mirror broken surfaces have the opposite mass. The [001] hinge states for strained SnTe is presented below, where two helical hinge states protected by mirror $(110)$ and $(1\bar{1}0)$ can be observed. \textbf{d} | The Bi crystal structure and Brillouin zone. The $(\bar{2}10)$ surface Brillouin zone is plotted, which corresponds to the surface band structures in \textbf{e}. The $C_2$ rotation is along the $[100]$ direction, which is the a-axis in the hexagonal plane. The mirror plane is normal to the $C_2$ rotation and denoted as $(\Bar{2}10)$ in the hexagonal coordinates. The $(\Bar{2}10)$ surface respects the $C_2$ rotation but breaks the mirror. The bulk band structure of Bi is presented on the right. \textbf{e} | The bulk Wilson loop and the $(\Bar{2}10)$ surface states. The Wilson loop is integrated along the [100] $C_2$ axis direction. The surface Brillouin is searched to locate both the connected Wilson loops and surface Dirac cones at generic k points. Schematic representation of $C_2$ related surface Dirac cones is shown in the inset. \textbf{f} | The $C_2$-protected hinge states (highlighted in the white circle) along the $[100]$ direction.}
    \label{fig:TCI}
\end{figure}

For a mirror TCI, if the nontrivial mirror plane is not normal to side surfaces but passes through the hinge, all the surface states are gapped out. However, helical hinge states arise in the mirror-projected lines (Fig.~\ref{fig:TCI}c). This type of TCI belongs to the HOTI phase \cite{hsieh2012topological,Zhang2013,song2017d,benalcazar2017quantized,Langbehn2017,schindler2018higher1}.
To demonstrate this scenario, we choose the [001] as the hinge direction and (100) and (010) as side surfaces. By using the tight-binding model with the uniaxial [110] strain, as suggested in REF.~\citeonline{hsieh2012topological,schindler2018higher1}, side surfaces are gapped out while two nontrivial mirror planes (110) and $(1 \Bar{1}0)$ survive, passing through two hinges. 
The Green's function method is applied, and the system is considered as finite in the [100] direction, half-infinite in the [010] direction and periodic in the [001] direction (the hinge direction). Two hinge states can be observed in Fig.~\ref{fig:TCI}c, protected by (110) and $(1\Bar{1}0)$ mirror planes. It also satisfies the relation $C_m = 2n$ for the mirror protected hinge with TRS, where $n$ is the number of Kramers pairs of hinge states and $C_m$ is the corresponding mirror Chern number~\cite{schindler2018higher1}.

Another TCI/HOTI example is the element Bi~\cite{schindler2018higher,hsu2019topology}. 
To reveal the band inversions, the parity eigenvalues at the TRIM points are shown in Fig.~\ref{fig:TCI}d. At $\Gamma$, all parity eigenvalues are ``+'' for the top three valence bands. In contrast, all other TRIM ($F, L, T$) have two ``--'' and one ``+''. Thus, Bi has double band inversions at the $\Gamma$ point~\cite{hsu2019topology}.
The crystal symmetries, including the inversion and the [100] $C_2$ rotation (along the $a$-axis in the hexagonal plane), define the topology of the double band inversion.
Unlike mirror Chern number, topological invariants associated with rotation symmetries are difficult to calculate~\cite{song2017d,fang2019new}. Fortunately, using symmetry indicators greatly simplifies this process. Bi possesses the nontrivial $\delta_{2^{100}}$, $\delta_{2^{100}_1}$ and $\delta_i$. We focus on the rotation invariant $\delta_{2^{100}}$, which dictates a pair of Dirac points on the ($\Bar{2}10$) surface with the $C_2$ symmetry. In this plane, two band inversions lead to two surface Dirac cones at the $\Gamma$ point. Two Dirac cones cross each other in the plane and gap out by hybridization except at two crossing points that are protected by $C_2$ . 
But these two points are not necessarily high-symmetric momenta in the surface Brillouin zone~\cite{fang2019new}.
Additionally, similar double band inversions also lead to the HOTI and related TSSs in $\beta$-MoTe$_2$ \cite{wang2019higher,tang2019efficient}, {which is protected by the $C_2$ screw rotation}. 

To identify the TSS of $\delta_{2^{100}}$, we cleave the $(\Bar{2}10)$ surface and search the surface Brillouin zone based on the tight-binding calculations. Two Dirac cones are found in Fig.~\ref{fig:TCI}e, and the corresponding Wilson loop shows the same topology.
The $\delta_{2^{100}}$ also dictates two helical hinge states on side hinges, which connect double Dirac cones on top and bottom. Unlike the mirror HOTI, the $C_2$ hinge states are unpinned and depend on the geometry of the sample \cite{fang2019new,hsu2019topology}.
Using the same method as that for SnTe (Fig.~\ref{fig:TCI}c), we calculated the hinge states in Fig.~\ref{fig:TCI}f, which are further compatible with the inversion invariant $\delta_i$~\cite{schindler2018higher} and the screw rotation invariant $\delta_{2_1^{100}}$. 
However, it is technically challenging to unambiguously address the hinge states of Bi~\cite{schindler2018higher,nayak2019resolving},  because neither the bulk nor side surfaces exhibit a global band gap, as indicated in Fig.~\ref{fig:TCI}f.

\subsection*{H2 Weyl Semimetals}

\begin{figure}[h!]
    \centering
    \includegraphics[width=\textwidth]{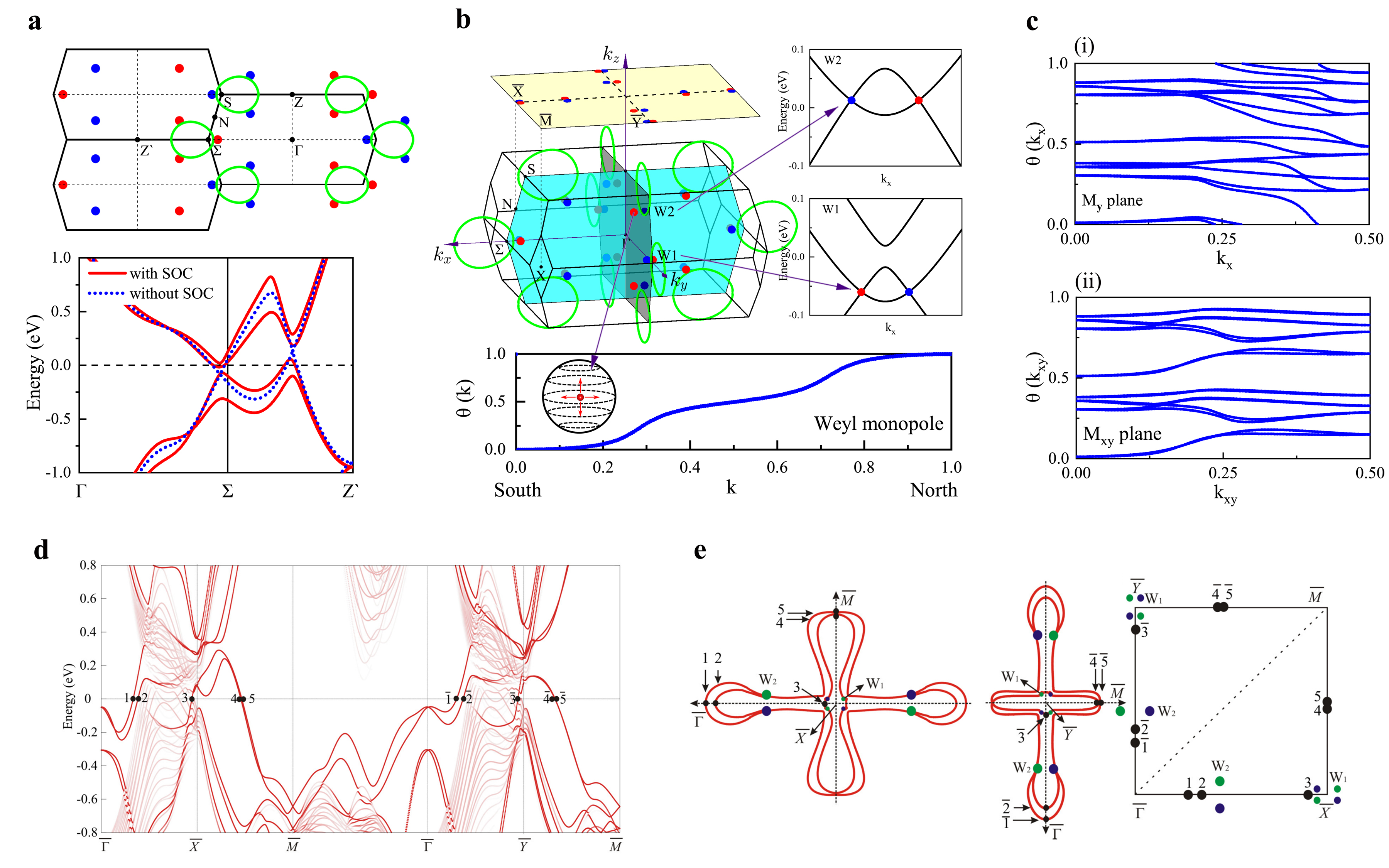}
    \caption{\textbf{Weyl Semimetals.} \textbf{a} | The nodal rings and $k$-path in the Brillouin zone of TaP (top). The band structure of TaP with and without spin-orbit coupling (SOC) (bottom).  \textbf{b} | Schematic representation of the nodal line structures (without SOC) and Weyl points (with SOC) in the Brillouin zone. The local band structure connecting a pair of W1 and W2 Weyl points is presented on the right. The Wilson loop of one W2 Weyl point is shown at the bottom, which is summed among all valence bands. \textbf{c} | The Wilson loop of the nontrivial mirror plane $M_y$ and the trivial glide mirror plane $M_{xy}$. \textbf{d} | The (001) surface band structure based on the first-principles density functional theory (DFT) calculations (with the P atom termination). \textbf{e} | The (001) surface projected Fermi surfaces around $\Bar{X}$ and $\Bar{Y}$. Black dots are crossing points in \textbf{d}. Red and blue points are projected Weyl points with opposite chiralities. Panel \textbf{d} and \textbf{e} are adapted with permission from REF.~\citeonline{sun2015topological}.}
    \label{fig:WSM}
\end{figure}

The WSM is a 3D topological semimetal phase without requiring any symmetry protection. Instead, it needs to break either the TRS or inversion symmetry (or both) to lift the spin degeneracy, because 
the coexistence of TRS and inversion symmetry forces the spin degeneracy at all $k$-points.
A WSM features two-fold-degenerate accidental-crossing points, called Weyl points, between conduction and valence bands.
Weyl points exhibit monopole-like Berry curvature distribution~\cite{fang2003anomalous} and always come in pairs with opposite chiralities. 
Because Weyl points occur at generic momenta, addressing Weyl points and induced Fermi arc surface states~\cite{wan2011topological} are technically challenging in calculations. 
In the following, we focus on WSM with TRS, where the symmetry indicator method usually cannot be applied (Fig.~\ref{fig:WSM}).

We take TaP as an example. TaP, with SG $\#$109, is a nonmagnetic WSM without inversion symmetry~\cite{huang2015weyl,weng2015weyl}. It contains two mirror planes $M_x$ and $M_y$, which are related to each other by the $C_4$ rotation. 
In the $M_y$ plane, the bulk band structure exhibits a clear inversion from $\Gamma$ to $\Sigma$, to $\mathrm{Z^\prime}$ in the nearby Brillouin zone (Fig.~\ref{fig:WSM}a). Without SOC, the band inversion induces two nodal rings in $M_y$ plane and another two in the $M_x$ plane in the first Brillouin zone. As SOC is turned on, we expect to find Weyl points near the nodal ring regimes, because other regimes are fully gapped. 
As shown in Fig.~\ref{fig:WSM}b, the nodal rings inside the $M_{x,y}$ planes are gapped out by SOC. Each original nodal ring evolves into three pairs of Weyl points near it, leading to twelve pairs of Weyl points in total. By searching for band crossing points near the nodal rings, Weyl points can be located. As long as one Weyl point is found at $(k_x,k_y,k_z)$, other Weyl points can be located at $(\pm k_x,\pm k_y, \pm k_z)$ and $(\pm k_y,\pm k_x, \pm k_z)$ by considering the $M_x$, $M_y$, $C_4$ and TRS.

In numerical calculations, it is usually difficult to distinguish a band crossing from a small finite gap. This problem can be solved by the monopole nature of Weyl points. 
Weyl points feature the source ($+$ chirality) and sink ($-$ chirality) of Berry curvature $\bm{\Omega(k)}$~\cite{xiao2010berry}, which can be calculated based on the Wannier-function Hamiltonian (see Eq.~\ref{Berry}). Thus, the Berry curvature flux for a closed surface surrounding a Weyl point is $\pm 2\pi$, where $\pm$ represents the chirality. 
To estimate the Berry flux, alternatively, the Wilson loop method is equivalent but more convenient than integrating the Berry flux. 
On the spherical surface embodying a Weyl point, the Wilson loop could be chosen as azimuthal rings. The Berry phase of valence bands evolves from the south pole to the north pole and presents the nontrivial Chern number $C=\pm1$, as shown in Fig.~\ref{fig:WSM}b.

The bulk Weyl points dictate the existence of surface Fermi arcs. However, when topological Fermi arcs and trivial surface states coexist, identifying Fermi arcs is challenging. For example, TaP has multiple pairs of Weyl points and thus multiple Fermi arcs connecting them. The surfaces are also terminated by P or Ta dangling bonds, leading to trivial states. Different surface terminations naturally exhibit different surface band structures that share the same topology. Furthermore, different computation methods (DFT and tight-binding) may give different surface states that should share the same topology \cite{weng2015weyl,lv2015experimental,xu2015discovery,yang2015weyl,sun2015topological}. 

Topological information of bulk mirror planes is helpful to understand surface state topology. The Wilson loops in FIG.~\ref{fig:WSM}c reveal their topological feature.  
Because of TRS, the $M_{x}$ and $M_y$ planes are actually 2D TIs.
In contrast, two diagonal glide mirror planes ($M_{xy}$ and $M_{\bar{x}y}$) are trivial 2D insulators, because the Wannier charge centers keep the same partners between $\Gamma$ and $\mathrm{M}$ points.

The $M_y$ and $M_{xy}$ planes are further projected to the surface $\Bar{\Gamma}$-$\Bar{X}$ and $\Bar{\Gamma}$-$\Bar{M}$ lines, respectively. The 2D TI of $M_y$ leads to odd pairs of helical edge states along the $\Bar{\Gamma}$-$\Bar{X}$ line.
In contrast, the trivial TI of $M_{xy}$ leads to even pairs of helical edge states along the $\Bar{\Gamma}$-$\Bar{M}$.
Therefore, there might be odd pairs of Fermi arcs crossing the $\Bar{\Gamma}$-$\Bar{X}$ but stopping somewhere before the $\Bar{\Gamma}$-$\Bar{M}$.
Indeed, FIG.~\ref{fig:WSM}d shows three surface states (point 1-3) between $\Bar{\Gamma}$ and ${\Bar{X}}$ but no state between $\Bar{\Gamma}$ and ${\Bar{M}}$ at the Fermi surface. 
In other words, the Fermi arcs can be considered as the extended edge states of the 2D TI (TRS-preserving) or the Chern insulator (TRS-breaking), as demonstrated in Fig.~\ref{fig:2D}. 

Since Fermi arcs connect Weyl points with opposite chiralities, we can also choose a closed loop in the surface Brillouin zone to identify the topology. If the closed loop contains the net odd chirality, there must be an odd number of Fermi cuts across the loop. Otherwise, there must be an even number of cuts. For instance, for the closed $\Bar{\Gamma}$-$\Bar{X}$-$\Bar{M}$-$\Bar{\Gamma}$ loop in FIG.~\ref{fig:WSM}e, it contains two positive W2 points (projected into the same point on the (001) surface) and one negative W1 point, with the net chirality $-1$. Fermi cuts across this loop are calculated to be five times (points 1-5 in FIG.~\ref{fig:WSM}d-e), as expected. The situation is also similar for the $\Bar{\Gamma}$-$\Bar{Y}$-$\Bar{M}$-$\Bar{\Gamma}$ loop.
If the surface boundary conditions are changed, the Fermi arcs may evolve dramatically~\cite{sun2015topological,Yang2019}.
Despite the complexity, the surface topology can still be identified by applying the bulk-boundary correspondence. 

WSMs have also attracted extensive studies in magneto-transport on the extremely large magnetoresistance~\cite{Shekhar2015} and the chiral anomaly effect~\cite{nielsen1983adler} characterized by the negative  magnetoresistance~\cite{xiong2015evidence,Huang2015,Zhang2016}. 
The 3D Fermi surface, especially energies of Weyl points with respect to the chemical potential, is essential to interpret experiments and exclude other effects like the current jetting~\cite{Arnold2016,dos2016search,Liang2018}.  
The Fermi surface can be directly calculated from the first-principles or related tight-binding calculations. 
The extreme cross-section ($A$) of the Fermi surface corresponds to the basic frequency ($F$) of the quantum
oscillations in experiments, by Onsager's relation 
$F=(\phi_0/2\pi^2)A$, where $\phi_0 = h/2e$ is the magnetic flux quantum. We can compare the angle-dependent quantum oscillations and the angle-dependent Fermi surface cross-sections, to determine the exact Fermi surface topology of WSMs \cite{Arnold2016b,Klotz2016,Arnold2016}. 

\subsection*{H2 Dirac Semimetals}
DSMs can be classified into band inversion-induced \cite{Wang2012,wang2013three} and symmetry-enforced DSMs \cite{young2012dirac}. The first type is unified in the band inversion scenario and can be viewed as two pairs of WSM: two Weyl points with opposite chiralities overlap and form the four-fold degenerate Dirac point. To maintain the band crossing, it requires the rotation symmetry ($C_n$, n = 3, 4 or 6) protection. Two Dirac points sit on the rotation axis, and they cannot be gapped due to different symmetry representations of inverted bands. 
Generally, to check the symmetry protected band crossings, we can calculate the symmetry representations of crossing bands from first-principles. If they belong to different representations, their hybridization is forbidden. 
To identify the topology of DSM, we can similarly calculate the $Z_2$ number of the 2D plane (for example, $k_z=0/0.5$) between Dirac points, in which the band gap is inverted. On the surface, the helical edge states of the 2D TI layer evolve into a pair of Fermi arcs. Two arcs meet each other at the projected Dirac points. 
The DSM can also transform into other phases by symmetry breaking. If a Dirac point picks up a mass term (opens a gap), the DSM becomes a TI or normal insulator. Otherwise, a Dirac point may split into two Weyl points if TRS or inversion symmetry is broken. 
The other type of DSM is enforced by the nonsymmorphic space symmetry. Such a space group has the four-dimensional irreps at some high-symmetry points of the Brillouin zone boundary. The symmetry requirement pins the Dirac points there and does not necessarily feature surface states. 

\subsection*{H2 Multifold-degenerate nodal points}

The symmetry-enforced Dirac points can be generalized to multifold-degenerate nodal points, where the three-, six-, and eight-fold band crossings may occur~\cite{Zaheer2013,wieder2016double,bradlyn2016beyond}. 
The multi-dimensional symmetry representations of nonsymmorphic space symmetries can admit such unconventional fermions in solids.
Since they cannot be described by the Dirac or Weyl equation, the general form of the Hamiltonian is formulated in REF.~\citeonline{bradlyn2016beyond} and further indicates that some band crossings possess the higher than one monopole charge~\cite{chang2017unconventional,tang2017multiple}. 

Triple-point semimetals are another type of semimetal with multifold band crossings, but are induced by the band inversion and protected by the $C_{3v}$ symmetry~\cite{Zaheer2013,weng2016topological1,zhu2016triple,weng2016coexistence,chang2017nexus,yang2017prediction}. 
Along the $C_{3v}$ rotation axis, the high-symmetry line admits 1D and 2D representations. When bands that belong to these two different representations are inverted, the crossing points on the rotation axis are triply degenerate and cannot be gapped out. 
Such a triple-point carries two monopole charges. It will split into two Weyl points with the same chirality if the rotation symmetry is slightly broken. When projected onto a surface, a pair of triple-points induces two Fermi arcs.  

\subsection*{H2 Nodal line semimetals}
In the 3D $k$-space, nodal points may connect each other and exhibit nodal lines. Nodal lines can form rings such as those in the TaP band structure without SOC (FIG.~\ref{fig:WSM}a) or form extend lines across neighboring Brillouin zones. Nodal lines without SOC are usually the precursor of other topological phases when including SOC. Related materials are called NLSMs.
With the TRS and inversion symmetry but no SOC, band inversion leads to the nodal line structure in the bulk and corresponding drumhead-like states on the surface~\cite{weng2015topological}, as shown in FIG.~\ref{fig:BI2}. 
However, the drumhead surface states are Shockley-type states and not necessarily robust against strong perturbations. 

The topology of NLSM can be identified by the Berry phase of a loop passing through the nodal line ($Z_2$ Berry phase NLSM) or the Wilson loop of the sphere containing the nodal line structure ($Z_2$ monopole NLSM) \cite{fang2015topological,fang2016topological,Ahn2018}. 
Heuristically, the single band inversion induces the former nodal line structure, and the double inversion may lead to the latter one~\cite{wang2019higher}.
For systems with mirror(glide) symmetry, the mirror(glide) planes can also host nodal lines if two bands with different mirror(glide) eigenvalues are inverted, like the above example of TaP without SOC. 
However, when switching on SOC, nodal lines can only be maintained in some special conditions, where TRS, inversion, and nonsymmorphic symmetry (such as glide) coexist to protect it \cite{fang2015topological,wieder2016spin,chen2015topological}. 

Except the WSM which does not require any symmetry protection, topological semimetals generally require TRS and/or crystal symmetries. Their band crossing points locate at high-symmetry points, lines or planes and can be probed by band representations. Therefore, Dirac points, multifold-degenerate nodal points, and nodal lines are incorporated into the symmetry indicator theory, and have been similarly identified in the topological materials databases from the symmetry perspective~\cite{vergniory2019complete,tang2019comprehensive,zhang2019catalogue}. 

\section*{ H1 Topology-induced Phenomena}

\subsection*{H2 Anomalous transport}

\begin{figure}[h!]
    \centering
    \includegraphics[width=\textwidth]{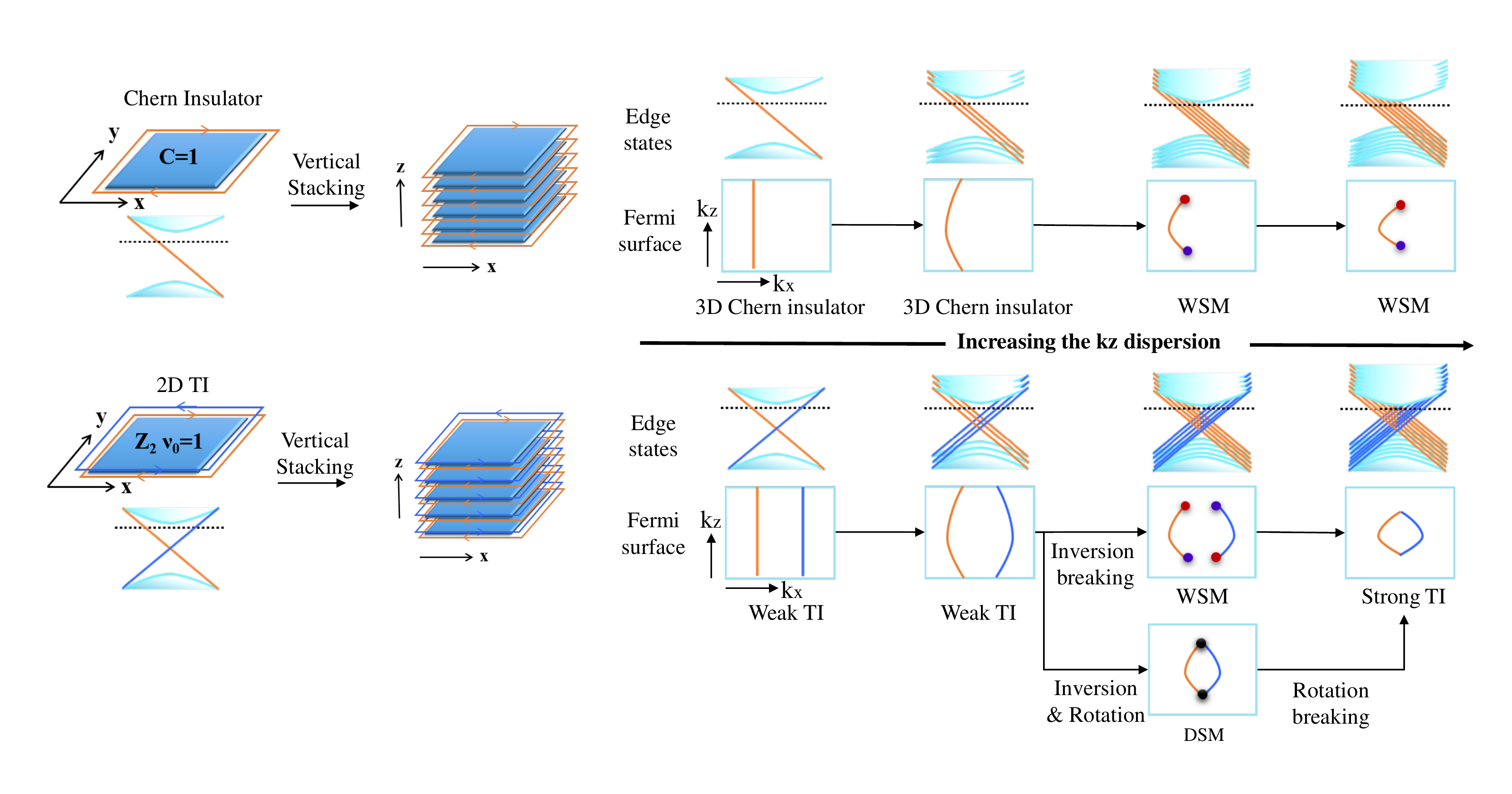}
     \caption{\textbf{3D topological phases.} 3D topological phases can be understood from the stacking of 2D Chern insulators or topological insulators (TI)s. The chiral/helical edge states of 2D layers form the surface states. Without the interlayer coupling, it belongs to a 3D Chern insulator/weak TI. Surface states appear on side surfaces, with no dispersion along $k_z$. With the increasing interlayer coupling, the $k_z$ dispersion is enhanced. 
Depending on the material, the band inversion in the 2D layers can be inverted, and  
3D topological states such as the weak TI, strong TI, Weyl semimetals (WSMs) and Dirac semimetals (DSMs) can emerge~\cite{Burkov2011a,weng2014transition}.
This layer construction scheme can also lead to topological crystalline insulators and higher order TIs ~\cite{song2018quantitative,yutinglayer}.
For example, the monolayer of MoTe$_2$ is a 2D TI~\cite{Qian2014QuantumSH} and the corresponding 3D material forms a WSM in the noncentrosymmetric phase ($T_d$)\cite{sun2015prediction,Wang2016} and a HOTI in the centrosymmetric phase~\cite{wang2019higher,tang2019efficient}}
  \label{fig:3D}
\end{figure}

Topological materials generally exhibit large SOC and large Berry curvature in the band structure. As demonstrated in FIG.~\ref{fig:3D}, we can understand the Fermi arcs of a WSM or DSM from the edge states of stacking the quantum anomalous Hall insulators or 2D TIs (quantum spin Hall insulators). It is natural to expect the anomalous Hall effect (AHE)~\cite{Nagaosa2010,weng2015quantum} in magnetic topological materials and the spin Hall effect (SHE)~\cite{Sinova2015} in nonmagnetic topological materials with strong SOC~\cite{Sun2016}.
From the Wannier function based tight-binding Hamiltonian ($H$), the Berry curvature and the anomalous Hall conductivity ($\sigma_{ij}^A$) can be calculated in the Kubo-formula scheme~\cite{xiao2010berry},
\begin{align}
    \Omega_{ij}^n(\mathbf{k}) &=2 \sum\limits_{m \neq n} Im[\frac{\langle u_{nk}|{dH}/{dk_j}| u_{mk} \rangle \langle u_{mk} | {dH}/{dk_i} | u_{nk} \rangle}{(E_m-E_n)^2}] 
    \label{Berry}
    \\
    \sigma_{ij}^A & = -\frac{e^2}{h}\sum\limits_{n} \int_{BZ} \frac{d^3 \mathbf{k}}{2\pi^3}f(E_n(\mathbf{k}))\Omega_{ij}^n(\mathbf{k})
    \label{sigma}
\end{align}
where $E_n$ and $| u_{nk} \rangle$ are the energy and Bloch wave function, respectively, $i,j=x,y,z$, and $f$ is the Fermi-Dirac distribution function. Equation~\ref{Berry} does not involve any differentiation on the wave function and thus can be evaluated under any gauge choice. It is particularly useful for numerical calculations. Usually an infinitesimal number is included in the denominator to avoid the division by zero. 

Magnetic WSMs and NLSMs indeed display strong AHE, as observed in Mn$_3$Sn and Mn$_3$Ge~\cite{Nakatsuji2015,Nayak2016,Yang2017}, Co$_3$Sn$_2$S$_2$~\cite{liu2018giant,wang2018large}, 
Co$_2$MnGa~\cite{Sakai2018} and Co$_2$MnAl~\cite{Li2020}. In these systems, the intrinsic, giant AHE usually dominates the anomalous transport. Thus, first-principles calculations and experiments usually obtain quantitatively consistent results. 
In contrast, the calculated SHE sometimes deviates from the experimental value, possibly because of the overwhelming extrinsic contributions. 
The strong AHE also indicates the large anomalous thermoelectric coupling (the Nernst effect) and the thermal Hall effect. The anomalous Nernst coefficient ($\alpha_{xy}^A$) can be estimated from the Mott relation~\cite{behnia2015fundamentals,Ding2019}, $\frac{\alpha_{xy}^A}{T} |_{T\rightarrow 0} = -\frac{\pi^2k_B^2}{3|e|} \frac{d\sigma^A_{xy}}{d\mu}$, where $\mu$ is the chemical potential. 
Here, ${\alpha_{xy}^A}$ comes from the Berry curvature on the Fermi surface, while $\sigma^A_{xy}$ corresponds to the Berry curvature summed over all the Fermi sea (Eq.~\ref{sigma}).
The thermal Hall conductivity ($\kappa^A_{xy}$) can also be estimated from the Wiedemann-Franz law~\cite{Xu2020}, 
$\frac{\kappa^A_{xy}}{T}|_{T\rightarrow 0} = L_0  \sigma^A_{xy}$, where 
$L_0=\frac{\pi^2}{3}(\frac{k_B}{e})^2$ is the Lorenz ratio. 

\subsection*{H2 Nonlinear optical phenomena}

Noncentrosymmetric topological materials like WSMs have attracted increasing interest in the nonlinear electric and optical phenomena, for example, nonlinear AHE in the presence of TRS~\cite{Deyo2009,Moore2010,Sodemann2015,Ma2019,Kang2019}, second harmonic generation~\cite{Wu2017} and
the generation of dc photocurrents (shift current and injection currents) under light irradiation~\cite{vonBaltz1981,Sipe2000,Young2012,Ma2017,osterhoudt2019colossal}.
Topological materials have triggered a revival of interest in the fundamental understanding of the photocurrent based on the quantum geometric phases (such as the Berry curvature and quantum metric) in the band structure~\cite{Hosur2011,Morimoto2016,Ventura2017,Parker2019,Holder2020}. 
In this direction, first-principles methods are being actively developed to investigate nonlinear phenomena in topological materials~\cite{Zhang2018a,Zhang2018b,Zhang2018,Facio2018,Zhang2019,wang2021giant,le2020ab,xu2020comprehensive,Juan2020,Chang2020,Fei2020,kaplan2020nonvanishing}.

\section*{H1 Useful Tools and Databases}

The topological classification of solid materials (especially nonmagnetic materials) is to some extent established with the assistance of first-principles calculations. From the material perspective, we are still searching for the perfect topological materials with negligible bulk carriers, for example, real-bulk-insulating TIs and the real-bulk-semimetallic WSMs that have all bulk Fermi pockets vanishing at the Weyl point. 

For new materials, the bulk topology can easily be evaluated based on the available calculations tools and databases (see TABLE 1). 
For instance, directly from DFT wave functions,  VASP2Trace~\cite{vergniory2019complete}, SymTopo~\cite{he2019symtopo}, or Irvsp~\cite{gao2020irvsp} can calculate band representation and identifiy symmetry indicator-based topological classifications. Wannier90~\cite{pizzi2020wannier90} is a powerful tool to extract maximally localized Wannier functions from DFT calculations. In the Wannier function basis, one can construct the tight-binding Hamiltonian and calculate the electronic, topological and transport properties. 
Based on tight-binding parameters from Wannier90, WannierTools~\cite{wu2018wanniertools} provides post-processing analysis. For instance, it can calculate the bulk topological invariants by the Wilson loop method, TSSs, and Berry phase-related properties. 

However, it usually requires more effort and attention to compute measurable quantities (such as TSSs and bulk quantum oscillations) to predict or interpret experiments.
Given that the topology refreshes our understanding of solid-state materials, there is an increasing interest in exploring the more significant consequences of topology in the transport and optical phenomena beyond measuring TSSs. To this goal, more computational methods and tools are being developed, based on the first-principles band structure and wave functions.

\bibliography{References}

\section*{Acknowledgements}
B.Y. acknowledges the financial support by the Willner Family Leadership Institute for the Weizmann Institute of Science,
the Benoziyo Endowment Fund for the Advancement of Science, 
Ruth and Herman Albert Scholars Program for New Scientists and the European Research Council (ERC) (ERC Consolidator Grant No. 815869, ``NonlinearTopo'').

\section*{Author contributions}
B.Y. conceived the review. J.X. made calculations and wrote the manuscript with inputs from B.Y.

\section*{Competing interests}
The authors declare no competing interests.

\section*{Peer Review Information} \textit{Nature Reviews Physics} thanks Quansheng Wu and other, anonymous reviewers for their contribution to the peer review of this work.

\section*{Supplementary information}
Supplementary information is available for this paper at https://doi.org/10.1038/s415XX-XXX-XXXX-X

\newpage 
\section*{Key points:} 
\begin{itemize}

\item Heuristically, the simple but intuitive band inversion scenario, which is easily accessible for first-principles calculations, can unify different topological states and rationalize their topological boundary states.

\item Quantitatively, topological invariants and symmetry indicators distinguish topological phases from atomic insulators. 

\item The Wilson loop reveals the bulk topology and also the  surface dispersion profile. 

\item The surface state topology is uniquely determined by the bulk state topology. But surface band dispersion changes as varying the specific surface condition.

\item The band structure topology leads to interesting anomalous transport and nonlinear optical phenomena.

\end{itemize}

\section*{Table}
 
\begin{table}[ht]
\caption{ Some useful methods, tools and databases for the first-principles calculations on topological quantum materials.}
\label{table1}
\begin{tabular}{cl}
\hline
Types of tools  & \multicolumn{1}{c}{Resources} \\ \hline
Topological materials databases & \href{http://materiae.iphy.ac.cn/}{Materiae~\cite{zhang2019catalogue,he2019symtopo}}, \href{https://topologicalquantumchemistry.org/#/}{Topological Materials Database~\cite{bradlyn2017topological,vergniory2019complete}},\\ &
\href{https://ccmp.nju.edu.cn/}{Topological Materials Arsenal}~\cite{tang2019comprehensive}... \\
Wannier function methods and tools & \href{http://www.wannier.org/}{Wannier90~\cite{pizzi2020wannier90}}, \href{http://www.wanniertools.com/}{WannierTools~\cite{wu2018wanniertools}}, \href{http://z2pack.ethz.ch/}{Z2Pack~\cite{gresch2017z2pack}}, \href{https://wannier-berri.org/}{WannierBerri~\cite{tsirkin2020high}}, ... \\
Band representations tools      & \href{https://www.cryst.ehu.es/}{Bilbao Crystallographic Server~\cite{aroyo2006bilbao1}}, \href{http://materiae.iphy.ac.cn/symtopo}{SymTopo~\cite{he2019symtopo}}, \href{https://www.cryst.ehu.es/cgi-bin/cryst/programs/topological.pl}{VASP2Trace~\cite{vergniory2019complete}}, 
\href{https://github.com/zjwang11/irvsp/}{Irvsp~\cite{gao2020irvsp}}, \href{https://github.com/stepan-tsirkin/irrep}{IrRep~\cite{iraola2020irrep}}, ... \\
Materials databases             & \href{http://www2.fiz-karlsruhe.de/icsd_home.html}{ICSD~\cite{hellenbrandt2004inorganic}}, \href{https://materials.springer.com/}{Springer Materials}, \href{https://materialsproject.org/}{Materials Project~\cite{jain2013commentary}}, ...\\
Visualization tools             & \href{https://jp-minerals.org/vesta/en/}{VESTA~\cite{momma2008vesta}}, \href{http://www.xcrysden.org/}{Xcrysden~\cite{kokalj1999xcrysden}}, ... \\ 
\hline
\end{tabular}
\end{table}

\section*{Figure captions}

\clearpage

\noindent \textbf{Key references:}\\
Ref.\citeonline{fu2007topological1} proposed the parity criteria and used them to predict topological insulator materials. \\
Refs. \citeonline{yu2011equivalent,Soluyanov2011,alexandradinata2014wilson} proposed the method of Wannier charge center evolution and Wilson loop to evaluate the topology. \\
Refs. \citeonline{po2017symmetry,bradlyn2017topological,song2018quantitative,khalaf2018symmetry,kruthoff2017topological} proposed the symmetry indicators to classify general topological states and materials. \\
Refs. \citeonline{zhang2009topological,hsieh2012topological,weng2015weyl} are well known examples for materials prediction. \\

\noindent \textbf{Glossary terms:}

\noindent\textbf{Space group}. A symmetry group that includes all crystal symmetries. There are 230 space groups in total for 3D crystals.

\noindent\textbf{Irreducible representation}. For a space group, a representation is a set of matrices, each of which responds to a symmetry operation. The relation of symmetry operations is equivalent to the calculation of matrices. The irreducible, block-diagonal form of the matrix representation is called irreducible representation.

\noindent\textbf{Wyckoff positions}. In a space group, Wyckoff positions denote the symmetry-allowed positions, including sites and multiplicity, where atoms can be found.
\\

\noindent \textbf{Website summary:} \\
First-principles calculations have been very successful in predicting topological quantum materials. This Technical Review covers topological band theory and provides a guide to the study of topological materials with the first-principles methods.

\end{document}